\documentclass[preprint,showpacs]{revtex4}
\usepackage{color,url,ulem,bm,amsmath,times,graphicx,epsfig}
\newcommand{\bmr}{\ensuremath{\bm{r}}}
\newcommand{\ei}{\ensuremath{\bm{e}_i}}

\newcommand{\bmu}{\ensuremath{\bm{u}}}
\newcommand{\tdiff}{\ensuremath{t_{\mathrm{diff}}}}
\newcommand{\cs}{\ensuremath{c_{\rm{s}}}}
\newcommand{\fieq}{\ensuremath{f_{i}^{\rm{eq}}}}
\newcommand{\myeq}[1]{Eq.\ (\ref{#1})}
\newcommand{\Myeq}[1]{Equation\ (\ref{#1})}
\newcommand{\myfig}[1]{Fig.~\ref{#1}}

\newcommand{\xc}{x_{\rm{c}}}
\newcommand{\tc}{t_{\rm{c}}}

\def\Peclet{P\'eclet}
\def\xc{\ensuremath{x_{\mathrm{c}}}}
\def\Pe{\ensuremath{\mathrm{Pe}}}

\newcommand{\hide}[1]{{}}
\newcommand{\strike}[1]{{}}

\begin{document}

% Title of the paper
\title{Effect of aspect ratio on transverse diffusive broadening: A lattice Boltzmann study}
%%% authors

\author{S.G. Ayodele$^1$, F. Varnik$^{1,2}$, and D. Raabe$^1$}
\affiliation{$^1$Max-Planck Institut f\"ur, Eisenforchung, 
Max-Planck Stra{\ss}e 1, 40237, D\"usseldorf, Germany.\\
$^2$Interdisciplinary Center for Advanced Materials Simulation,
Ruhr University Bochum, Stiepeler Stra{\ss}e 129,  44780 Bochum, Germany.}

\begin{abstract}
We study scaling laws characterizing the inter-diffusive zone between two miscible fluids flowing side by side in a Y-shape laminar micromixer using the lattice Boltzmann method. The lattice Boltzmann method solves the coupled 3D hydrodynamics and mass transfer equations and incorporates intrinsic features of 3D flows related to this problem. We observe the different power law regimes occurring at the center of the channel and close to the top/bottom wall. The extent of the inter-diffusive zone scales as square root of the axial distance at the center of the channel. At the top/bottom wall, we find an exponent 1/3 at early stages of mixing as observed in the experiments of Ismagilov and coworkers [Appl. Phys. Lett. {\bf 76}, 2376 (2000)]. At a larger distance from the entrance, the scaling exponent close to the walls changes to 1/2 [J.-B. Salmon et al J. Appl. Phys. {\bf 101}, 074902 (2007)]. Here, we focus on the effect of {\it finite} aspect ratio on diffusive broadening. Interestingly, we find the same scaling laws regardless of the channel's aspect ratio. However, the point at which the exponent 1/3 characterizing the broadening at the top/bottom wall reverts to the normal diffusive behavior downstream strongly depends on the aspect ratio. We propose an interpretation of this observation in terms of shear rate at the {\it side} walls. A criterion for the range of aspect ratios with non-negligible effect on diffusive broadening is also provided.
\end{abstract}
\maketitle

\section{INTRODUCTION}
Microfluidic devices are becoming a means of performing low cost and high-throughput chemical and biochemical analysis on chip. The range of applications include measuring dynamics of protein folding~\cite{Lipman2003}, kinetics of enzyme reactions~\cite{Seong2003} and surface patterning of cells and proteins~\cite{David2002}. Some of these applications generally involve cross-stream interaction of two or more fluids flowing side by side in a channel. However, such systems are characterized by laminar flow due to the small dimensions involved and thus fluids flowing side by side can only mix or interact by molecular diffusion.

In pressure driven flow through rectangular microchannels, where the fluid motion through the channel is actuated by pressure pumps, the channel velocity profile is approximately parabolic across the shortest dimension, as dictated by the balance between the gradient of the applied pressure and the viscous shear stress in combination with the no-slip boundary condition at the walls. Due to slow fluid motion close to the walls, tracer particles are hardly advected by the flow, thereby spending longer time at the walls. At the center of the channel, on the other hand, the flow velocity is quite high and tracer particles are transported more efficiently. This results in a fairly wide distribution of the amount of time spent by tracer particles in the channel, the so-called residence time. As demonstrated  experimentally~\cite{Kalmholz1999,Ismagilov2000}, this wide distribution of residence times gives rise to a non-uniform diffusive broadening across the channel. Ismagilov et al.~\cite{Ismagilov2000}, for example, showed using confocal fluorescence microscopy that in contrast to the center line of the channel, where the extent (along $y$ direction) of the interdiffusion zone exhibits normal diffusive behavior $\delta \sim x^{1/2}$ ($x$=distance from the inlet), it scales as one third power of the axial distance, $\delta \sim x^{1/3}$, close to the top and bottom walls. These observations are shown to be in line with results of a scaling analysis, which makes use of a certain similarity with the general L\'{e}v\^{e}que problem ~\cite{Leveque1928,Ismagilov2000}.

Numerical solution of the advection-diffusion problem~\cite{Kalmholz2002,Salmon2007,Jimenez2005} confirms the existence of the experimentally observed scaling behavior with an exponent of 1/3 in the proximity of the walls as compared to the exponent 1/2 at the center of the channel. Furthermore, these calculations also show that the exponent 1/3 can only be observed if the distance (along the flow) from the entrance of the channel is not too large. At sufficiently large distances from the inlet, on the other hand, the exponent for diffusive broadening close to the walls approaches 1/2 and thus becomes identical to the scaling exponent in the center of the channel. The cross over distance, $\xc$, is identified as the distance at which the tracer concentration along the shortest channel dimension ($z$ direction) becomes homogeneous (note that $\delta$ is measured along the $y$ direction).

The above mentioned numerical approaches  assume a one dimensional parabolic velocity profile and neglect any dependence of the fluid velocity on the distance $x$ from the inlet as well as on the ``neutral'' direction, $y$. The first assumption is valid at axial distances $x \ge W, H$ ($W$=width and $H$=height of the channel), where the flow is fully developed. The dependence on $y$, on the other hand, can only be neglected if the width of the channel is large compared to the height $H$, i.e.\ in the case of large aspect ratio, $W/H\gg 1$. In the experiments~\cite{Ismagilov2000}, however, the fully developed fluid velocity is {\it two} dimensional, i.e.\ it depends both on $y$ and $z$ due to the finite aspect ratios investigated ($W/H=2-5$)~\cite{Gondret1997,Zheng2008}. Despite this fact, the agreement between theory and experiment is quite good suggesting that the specific form of the velocity profile does not play a crucial role as long as a linear regime close to the walls and an approximately uniform flow at the channel center can be assumed.

\begin{figure}
\centering
\includegraphics[scale=0.55]{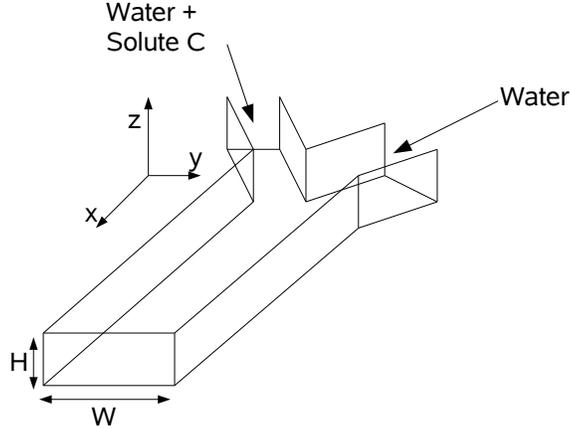}
\caption{Geometry of the micromixer with two inlets at $45^{o}$. The origin of the co-ordinate system is at the middle of the entrance. Measurements are taken after reaching a fully developed flow}
\label{fig:Y-junction}
\end{figure}

In the present work, we are going to focus on this aspect via a systematic study of the effect of finite aspect ratio on the spreading dynamics of a tracer field entering the Y-junction micromixer through one of the arms  (see Fig.~\ref{fig:Y-junction}). For this purpose, we solve, via the lattice Boltzmann (LB) method \cite{Qian1992}, the full three dimensional advection-diffusion problem for rectangular channels with various aspect ratios. Our previous studies of wall roughness effects on the chaotic mixing of passive tracers in a 2D channel showed the flexibility of the LB method in dealing with advection-diffusion problem in microchannels~\cite{Varnik2006,Varnik2007a,Varnik2007b}. The present work extends this approach to 3D with a particular focus on various scaling laws within the {\it laminar} flow regime using smooth walls (no wall roughness effects).

In addition to a systematic study of the effects related to a finite aspect ratio, our studies differ from numerical calculations of Salmon and Ajdari in that we do not make any assumption about the shape of the velocity profile. Rather, the velocity profile results from the solution of the Navier-Stokes equations for the problem of interest. By doing this, we remain as close as possible to real experiments and realize a range of velocity profiles from practically parabolic to those with strong deviations from parabolic dependence.

In the following section, we briefly introduce the simulation scheme and also provide some benchmark tests for our LB simulation by comparing our results with known analytical solutions for the spreading of a point source in a 3D microchannel both with and without walls. Excellent agreement with the analytical solutions is found. We then apply in section \ref{sec:results} the LB method to study the extent of the inter-diffusion zone between two fluids flowing side by side in a Y-shape micromixer. We study the combined effect of \Peclet{} number (defined as the ratio $\Pe=UH/D$, where $U$ is the maximum fluid velocity, $H$ the channel height and $D$ the diffusion coefficient of the tracer field) and channel's aspect ratio on the broadening. Our studies include both the upstream zone where the non-uniform broadening along the $y$ direction occurs (with an extent $\delta \sim x^{1/3}$ close to the walls and $\delta \sim x^{1/2}$ at the center of the channel) and downstream region where diffusion has had enough time to homogenize the concentration distribution along the vertical $z$ direction whereby leading to homogeneous broadening with a one-half power law both at the walls and in the center of the channel. A summary compiles our results.

\section{NUMERICAL METHOD AND ITS VALIDATION}
\label{sec:method}

\subsection{The Lattice Boltzmann Method}
The lattice Boltzmann method~\cite{McNamara88,Higuera89,Qian92,succi2001} is a mesoscopic approach that approximately solves the continuum Boltzmann equation in a discrete form by dividing space into a regular lattice, time into discrete steps and velocities into a finite number of vectors~\cite{He1997}. The number and direction of the velocities are chosen such that the resulting lattice is symmetric so as to easily reproduce the isotropy of the fluid~\cite{Rubinstein2008}. The density of the fluid at each lattice site is accounted for by a one particle probability distribution $f_{i}(\bmr,t)$, where the subscript $i$ represents one of the finite velocity vectors $\ei$, $\bmr$  the lattice site and $t$ the time. During each time step particles stream along each velocity vector $\bm{e_{i}}$ to a neighboring lattice site  and collide locally, conserving mass and momentum in the process. The LB equation describing propagation and collision of the particles is given by:
\begin{equation}
 f_{i}(\bmr+\ei, t+1)-f_{i}(\bmr, t)=\Omega_{i}f_{i}(\bmr, t),
\label{eq:fi}
\end{equation}
  where $\Omega_{i}$ is the collision operator.

The most widely used LB method is the lattice BGK model~\cite{Qian1992} which approximates the collision operator by simplifying it to a single time relaxation towards the local equilibrium distribution $f^{eq}_{i}$. The lattice BGK model is given as:
\begin{equation}
 f_{i}(\bmr+\ei, t+1)-f_{i}(\bmr, t)=\frac{\fieq(\bmr, t)-f_{i}(\bmr, t)}{\tau}
\label{eq:relax}
\end{equation}
 where $\tau$ is relaxation time and the equilibrium distribution $\fieq$ is closely related to the low Mach number expansion of the Maxwellian velocity distribution given as~\cite{Frisch1987}
\begin{equation}
 \fieq(\bmr, t)= w_{i}\rho\left[ 1 + \frac{1}{\cs^{2}}(\ei \cdot \bmu) + \frac{1}{2\cs^4}(\ei \cdot \bmu)^{2}-\frac{1}{2\cs^{2}}u^{2}\right] 
\label{eq:Maxwellian}
\end{equation}
 In~\myeq{eq:Maxwellian}, $\cs$ is the sound speed on the lattice and $w_{i}$ is a set of weights normalized to unity. Near equilibrium and in the limit of small Knudsen number (= mean free path/characteristic length of problem) the macroscopic Navier Stokes equation can be recovered using Chapman-Enskog multiscale analysis~\cite{Koelman1991}. The relaxation time $\tau$ is then found to be related to the kinematic viscosity as  $\nu=(2\tau-1)/6$.
%\begin{equation}
% \nu=\frac{2\tau-1}{6} 
%\label{eq:Viscosity}
%\end{equation}

The density $\rho$, velocity $\bmu$ and pressure $p$ are computed from the distribution function using:
\begin{equation}
 \rho=\sum_{i=0}^Nf_{i} ; \qquad  \qquad 
 \rho \bmu=\sum_{i=0}^Nf_{i}\ei ; \qquad \qquad
  p=\rho \cs^{2}
 \label{eq:rho}
\end{equation}
 The weights $w_{i}$ in the equilibrium distribution depend on the number of velocities used for the lattice. In this work we use the D3Q15 model which has fifteen velocities with weights $w_{i}$ given as:
\begin{equation}
 w_{i} = \left\{ \begin{array}{ll}
         2/9 & \mbox{$\ei =(0,0,0),  i=0$};\\
         1/9 & \mbox{$\ei =(\pm1,0,0),(0,\pm1,0),(0,0,\pm1)\;\;  i=1....6$};\\
         1/72 & \mbox{$\ei =(\pm1,\pm1,\pm1),  i=7....14$} \end{array} \right.
\label{eq:Weights}
\end{equation}

The approach can be extended to simulate diffusing solute by introducing passive tracers into the flow field. The tracers are advected with the flow velocity $\bmu$ and follow similar relaxation and propagation as the fluid medium. In this study, the tracer field ${g_i}$ does not have any effect on the velocity field (passive tracer limit). The LB equation governing propagation and collision of the density (concentration) distribution of the passive tracers is given as:
\begin{equation}
  g_{i}(\bmr + \ei, t+1)-g_{i}(\bmr, t)=\frac{g_{i}^{eq}(\bmr, t)-g_{i}(\bmr, t)}{\tau_{d}}\qquad,
  \label{eq:relax_tracer}
\end{equation}
with the equilibrium distribution function, $g_{i}^{eq}(\bmr, t)= w_{i}C\left[ 1 + (\ei \cdot \bmu) / \cs^2 \right]$. Here, $C$ is the concentration distribution of the tracers given as $C=\sum_{i=0}^Ng_{i}$.

Using Chapman-Enskog multiscale analysis, the advection-diffusion equation can also be recovered in the limit of low Mach number and near equilibrium situation. The relaxation time $\tau_{d}$ is then found to be related to the tracer diffusion coefficient as $D=(2\tau_{d}-1)/6$.

\subsection{The Y-shape laminar micromixer}
In the Stokes flow regime, assuming a steady flow with a constant pressure gradient $\partial p / \partial x$ along the channel, the Navier-Stokes equation can be simplified as $ \partial p / \partial x =\eta \nabla^{2}u_{x}(x,y,z)$. For a rectangular channel of length $L$, width $W$ and height $H$ with a fully developed flow $\left( u_{x} = u_{x}(y,z)\right)$, this equation can be solved via the method of separation of variables with no-slip boundary condition on the walls of the channel. One thus obtains the following result for the velocity field~\cite{Schlichting1958}:

\begin{equation}
u_{x}(y,z) = \frac{ \partial p / \partial x}{2\eta L}\left[\left(\frac{H^{2}}{4}-z^{2}\right)             -\frac{8H^{2}}{\pi^{3}}\sum_{n=1,3,5}^\infty\frac{1}{n^{3}}\frac{\cosh\left( \frac{n\pi y}{H}\right) }{\cosh\left( \frac{n\pi W}{2H}\right) }\sin\left( \frac{n\pi}{H}\left(z+\frac{H}{2}\right) \right) \right].
\label{eq:velocity_profile}
\end{equation}

Far away from the side walls and close to the center of the channel $\left( y \approx 0 \right)$,  the fluid velocity follows the well known 2D parabolic profile. In the Y-shape micromixer studied here, the velocity profile at the two inlets is quite different from that in the mixing channel. Assuming that the channel width $W$ is larger than its height, $W>H$, it takes approximately a distance of the order of $W$ from the junction for velocity profile to become fully developed.

In the steady state, the mass transport of a solute of concentration $C$ with a constant isotropic diffusion coefficient $D$ in a unidirectional and one dimensional velocity field, $\bmu=({u}_x(z), 0, 0)$, can be describe by 

\begin{equation}
 {u}_{x}(z)\frac{\partial C}{\partial x}=D\left( \frac{\partial^{2}}{\partial x^{2}}+\frac{\partial^{2}}{\partial y^{2}}+\frac{\partial^{2}}{\partial z^{2}}\right)C.
\label{eq:mass_transfer}
\end{equation}

Given that the initial concentration at the junction, where the two fluids meet, is $C(x=0)=C_{0}$ and the characteristic fluid velocity is $U$, \myeq{eq:mass_transfer} can be put into a non-dimensional form using relevant characteristic length scale. A convenient form is to use $x^{*}=x/(\Pe H),\; y^{*}=y/H$ and $z^{*}=z/H$. \Myeq{eq:mass_transfer}  then becomes

\begin{equation}
 {u}_{x}(z^{*})\frac{\partial C}{\partial x^{*}}= \left( \frac{1}{Pe^{2}}\frac{\partial^{2}}{\partial {x^{*}}^{2}}+\frac{\partial^{2}}{\partial {y^{*}}^{2}}+\frac{\partial^{2}}{\partial {z^{*}}^{2}}\right)C.
\label{eq:non-dimensional}
\end{equation}

The first term on the right hand side of \myeq{eq:non-dimensional} represents the contribution of axial diffusion.  It can be neglected if {\it two} conditions are fulfilled. The first condition is that the  first term on the right hand side must be small compared to the other two terms i.e.\ $\partial^{2}C/\partial {x^{*}}^{2}/\Pe^{2}  \ll  \partial^{2}C/\partial {y^{*}}^{2},  \partial^{2}C/\partial {z^{*}}^{2}$. Obviously, this happens when $\Pe \gg 1$.  The second condition is that the axial convective term must be large compared to the axial diffusive term, i.e.\ $ {u}_{x}(z^{*})\partial C/\partial x^{*} \gg (\partial^{2}C/\partial {x^{*}}^{2})/\Pe^{2} $. In this case, using order of magnitude analysis, the regimes ($x^{*},z^{*}$) where this condition is valid can easily be identified to be~\cite{Salmon2007}, $x^{*} \ll 1/(\Pe^2u_x(z^{*}))$.

Assuming the validity of both the conditions stated above, \myeq{eq:non-dimensional}  can be simplified to 
\begin{equation}
{u}_{x}(z^{*})\frac{\partial C}{\partial x^{*}}= \left(\frac{\partial^{2}}{\partial {y^{*}}^{2}}+\frac{\partial^{2}}{\partial {z^{*}}^{2}}\right)C
\label{eq:reduced_eq}
\end{equation}

Salmon and Ajdari~\cite{Salmon2007} performed a numerical integration of \myeq{eq:reduced_eq} assuming a parabolic velocity profile across the channel, i.e. $u_x=u_x(0)[1-(2z/H)^2]$. The lattice Boltzmann method, on the other hand, solves the full \myeq{eq:mass_transfer} incorporating  all relevant 3D features such as the fact that, in general, the fluid velocity is not parabolic and depends on all the three coordinates $x, y$ and $z$, i.e. $u_x=u_x(x,y,z)$ \footnote{The dependence on $x$, however, can be neglected after a length of the order of $W$, necessary for the establishment of a fully developed flow profile across the channel.}. 

\subsection{Simulation details}
    
All simulations reported in this work are conducted using the D3Q15 LB model on a parallel machine. The lattice Boltzmann code is written in c++ with MPI (message passing interface) routines that enable exchange of information between processes. The code is run on SMP 8 x 2 core opteron 2.4 GHz. Typical channel dimensions used vary in the range $40 \times 20\times 1000$ to $400\times 20\times 1000$ as well as a channel of size $40 \times 40\times 1000$ (lattice units), thereby allowing the study of aspect ratios between 1 and 20. At the walls of the channel, the bounce-back scheme~\cite{succi2001} is implemented. Populations streamed to the walls are simply reversed back along the directions where they came from. Starting with the fluid at rest, a flow is imposed by the application of a body force. Note that it takes a finite amount of time (of the order of the momentum diffusion time $\tdiff=H^2/(8\nu)$~\cite{Varnik2007a}) until steady state is reached with respect to the fluid velocity, i.e.\  until the velocity field becomes independent of time. In order to avoid this transient effects, we wait a time of $5 \times \tdiff$ before injecting the tracer field.

After a stationary flow has been reached, we start to continuously inject passive tracers into the channel. This is simply achieved by setting at each time step the concentration of the tracers at the inlet to 1 (i.e.\ $C(x=0; t) \equiv C_0=1$ for all times $t$). We then monitor the time evolution of the tracer concentration field along the channel. Computation of the extent of diffusive broadening of the concentration distribution at a given cross section is done after a time independent concentration profile has been reached at the area of interest (corresponding to a dynamic balance between the incoming and outgoing tracer populations). The modeled diffusing analyte in this work has a diffusion coefficient of $D=0.5 \times 10^{-9}$m$^2$/s corresponding to $10^{-4}\Delta x^{2}/\Delta t$, where we have chosen the lattice units of $\Delta x=3\mu$m and $\Delta t= 600$ns. The kinematic viscosity  of the fluid was set to $\nu=0.1\Delta x^{2}/\Delta t$ corresponding to the viscosity of water ($\nu_{\mathrm{water}} \sim 10^{-6}$m$^{2}$/s at room temperature). The parameters chosen are readily comparable with that of the experiments.\\

\subsection{A point source in a 3D channel}
Before presenting in section \ref{sec:results} results of our simulations on laminar micromixer, we provide here a simple test of our lattice Boltzmann approach. For this purpose, we compute the fundamental problem of diffusion of a point source placed at point $\bmr_0=(x_{0}, y_{0}, z_{0})$ at time $t=0$ on a 3D square lattice using the LBGK model. Assuming a constant and isotropic diffusion coefficient $D$, the system is described by the well known diffusion equation $\partial C/\partial t=D\Delta C$ ($\Delta$=the Laplace operator) with the initial condition $C(\bm{r}, 0)=\delta(x-x_0) \delta(y-y_0) \delta(z-z_0)$. The fundamental solution or Green function of this equation is well known to be~\cite{Mathews1970}
\begin{equation}
 G(\bmr,\bmr_0, t)=\frac{1}{(4\pi Dt)^{3/2}}
 \exp\Big(-\frac{(x-x_0)^{2}-(y-y_0)^{2}-(z-z_0)^{2}}{4Dt}\Big).
\label{eq:Green_function}
\end{equation}

\begin{figure}
\centering
\includegraphics[scale=0.6]{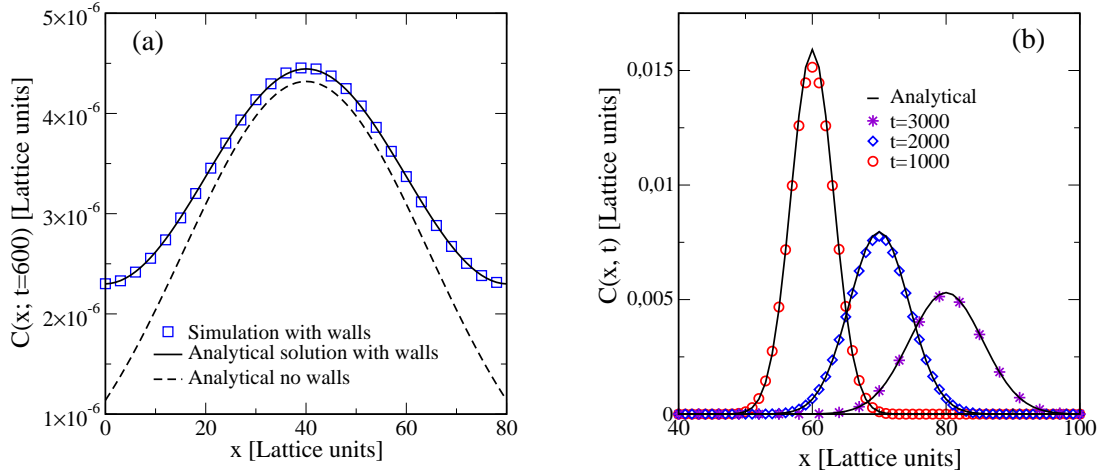}
\caption{Comparison of lattice Boltzmann simulation with analytical results (a) diffusion of a point source within a cubic box surrounded by non-absorbing walls (equivalent to a no flux boundary condition). (b) diffusion of a single point source in a fluid moving with a uniform velocity. The dashed line in (a) represents the solution of the same problem in an infinite system (no walls) thus underlying the non-trivial effect of the walls, fairly well captured by the LB method.}
\label{fig:LB_and_analytical}
\end{figure}

In the presence of non-absorbing impermeable walls (equivalent to no-flux boundary condition), the analytical solution for the diffusion of a point source can be obtained by using the principle of superposition and method of images~\cite{Crank1956}. The positions $(x_{k}, y_{m}, z_{n})$ of the infinite number of images  formed on each side of the box is obtained from the real point source according to the relations $x_{k} = x_{0}+ k L_{x}$,  $y_{m} = y_{0}+ m L_{y}$ and $z_{n} = z_{0}+ n L_{z}$. The final result for the concentration profile in the presence of non-absorbing walls is thus
\begin{equation}
 C(\bmr, t)= \frac{1}{(4\pi Dt)^{3/2}}\sum_{k,m,n=-\infty}^\infty \exp \Big(-\frac{(x-x_{k})^{2}-(y-y_{m})^{2}-(z-z_{n})^{2}}{4Dt} \Big).
\label{eq:solution}
\end{equation}
Using a square lattice of dimension $L_{x}$=80, $L_{y}$=80, $L_{z}$=80 and diffusion coefficient $D=0.5$ (lattice units), we perform lattice Boltzmann simulation with a simple bounce back scheme~\cite{succi2001} at the walls. Figure \ref{fig:LB_and_analytical}(a) compares the  simulation results for the concentration profile along the $x$-direction with the analytical solution obtained from \myeq{eq:solution}. In the same figure we also plot the Gaussian function given by \myeq{eq:Green_function}, which represents the solution of the same problem in the absence of the walls, thus emphasizing the non-triviality of the obtained solution in the presence of the walls. Our simulation results are in agreement with the analytical results obtained from \myeq{eq:solution} within an error of 0.2\%. This agreement shows that LB model is able to capture the effect of the wall correctly. It is straight forward to extend the situation to the case of diffusion in a uniform steady unidirectional velocity field $\bm{u}=(U_{0}, 0, 0)$ . The results for this situation are shown in Fig.~\ref{fig:LB_and_analytical}(b) for the case of an infinite system (no walls). Here, the point source exhibits a normal diffusive broadening ($\delta \sim t^{1/2}$), while at the same time being advected by the flow. The analytic curves shown in \myfig{fig:LB_and_analytical}(b) are flow advected version of Gaussian distributions given in \myeq{eq:Green_function}, where $x-x_0$ is replaced by $x-(x_0+ U_0 t)$. Noting that the center of the Gaussian distribution along the flow direction obeys $\left< x \right>=U_0t$, the normal diffusive broadening can also be expressed as $\delta \propto t^{1/2} \propto \left< x \right>^{1/2}$.

\section{Results and Discussion}
\label{sec:results}
\begin{figure}
\centering
\includegraphics[scale=0.95]{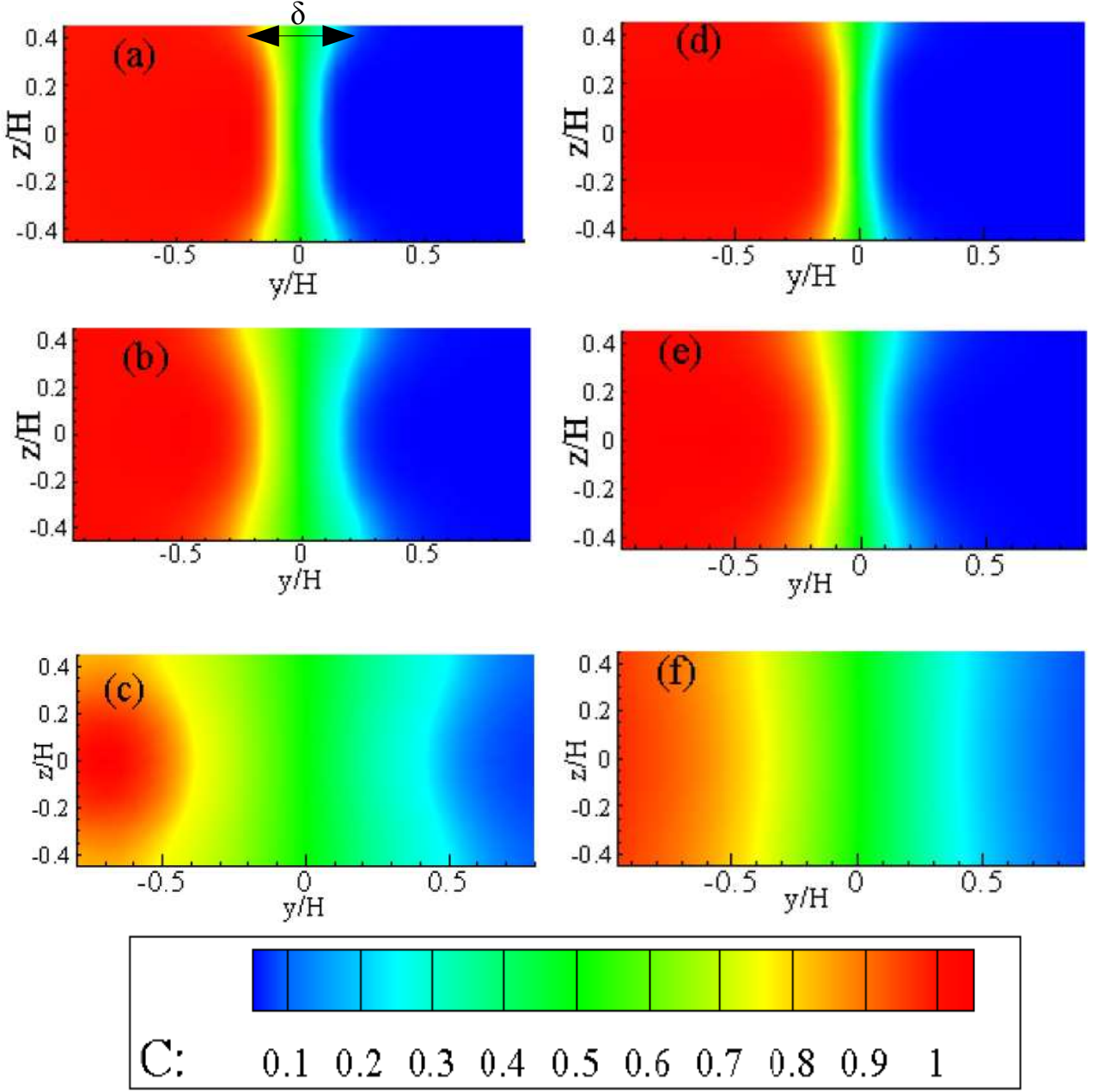}
\caption{Solute concentration $C$ in $y-z$ cross sections of a rectangular channel for aspect ratios $W/H=2$ (left panels) and 5 (right panels) at various reduced distances $x^*$ from the inlet. From top to bottom:  $x^{*}=10^{-2.6},\;\;10^{-2.3}$ and $10^{-1.12}$. For all cases shown, the \Peclet{} number is $\Pe=1000$ and  the height of the channel is $H=20$ lattice units. The width of the channel is $W=40$ (left panels) and $W=100$ (right panels) lattice units.}
\label{fig:Images}
\end{figure}

For a \Peclet{} number of $\Pe=1000$, we plot in \myfig{fig:Images}, the cross-sectional image of the concentration field at various axial positions $x^*=x/(H\Pe)$ along the channel for aspect ratios of $W/H=2$ and 5 respectively. First note that the width of the interdiffusion zone increases with increasing axial distance (compare panels (a), (b) and (c)). This is reminiscent of the role played by time in diffusion process. Indeed, recalling that the tracer concentration at a height $z$ is advected with the velocity $u(z)$ and neglecting the effect of shear for the moment, a rough correspondence between diffusion time and the distance $x$ from the inlet can be obtained via $t(x,z)=x/u(z)$. A larger distance from the inlet thus corresponds to a larger diffusion time.

A survey of  \myfig{fig:Images} allows a second important observation, namely that the extent $\delta$ of interdiffusion zone increases when going from the center of the channel towards the walls, the so called ``butterfly effect'' (see e.g.\ panel (b)). Again, making use of the above estimate of diffusion time  as $t(x,z)=x/u(z)$, a qualitative understanding of this behavior can be gained by noting that $u(z)$ is maximum at the center of the channel ($z=0$) and decreases to zero at the walls ($z=\pm H/2$). Thus, at a given distance $x$ from the inlet, the time available for diffusion is larger in the proximity of the walls as compared to the center of the channel.

Even though being able to describe some important qualitative features of diffusive broadening, the above argument is too crude to capture the different scaling laws discussed above ($\delta \propto x^{1/3}$ close to the walls as compared to $\delta \propto x^{1/2}$ at the channel center). Indeed, an argument based only on an estimate of the effective diffusion time would yield $\delta=\sqrt{6Dt}=\sqrt{6Dx/u(z)}\propto x^{1/2}$. In other words, one would obtain diffusive broadening with an exponent of $1/2$ for all distances from the wall. the $z$-dependence being reflected in an effective diffusion coefficient $D/u(z)$. An adequate description of experimental observation, therefore, requires taking into account the effect of the non-uniformity of the flow on tracer distribution. Interestingly, assuming a linear velocity profile seems to be sufficient for a derivation of the observed scaling exponent of $1/3$ close to the walls~\cite{Ismagilov2000,Salmon2007}. The exponent $1/2$, on the other hand, results from the presence of a quasi-uniform flow in the central region of the channel.

A third important aspect in \myfig{fig:Images} is the effect of aspect ratio. A comparison of the left and right panels of \myfig{fig:Images} at a given distance from the inlet clearly shows that the interdiffusion zone is broader in the case of lower aspect ratio. As a consequence, also the inhomogeneity of the tracer concentration field along the $z$ direction is more pronounced when the aspect ratio is decreased. One could therefore expect that a longer diffusion time is required in order to homogenize the concentration field across the vertical direction.  Using the rough correspondence between time and axial distance along the lines discussed above, the position at which the concentration distribution across $z$ direction becomes homogeneous should increase for smaller aspect ratios. A comparison of panels (c) and (f) in \myfig{fig:Images} confirms this expectation. Both these panels correspond to the same distance from the inlet but different aspect ratios. While in \myfig{fig:Images}(f) (aspect ratio=5) the solute concentration is already homogeneous along the $z$-direction, it exhibits significant inhomogeneity in \myfig{fig:Images}(c) (aspect ratio=2). An important consequence of this feature will be worked out below.

In order to proceed with a more quantitative analysis, we adopt a simple definition of the extent of  interdiffusion zone  $\delta$  as the width of the concentration profile at which the concentration is reduced to 20\% of the maximum value at the inlet. It is to be stressed that the results presented here are insensitive to the specific definition of $\delta$. Other definitions such as using the integral over the second moment of the spatial derivative of the concentration field, $\delta= (\int dy y^2 \partial C/\partial y )/(\int dy \partial C/\partial y )$, lead essentially to the same conclusions. The present simple definition, however, is numerically more robust since it does not require the computation of numerical derivative of the concentration field at discrete intervals.

\begin{figure}
\centering
\includegraphics[scale=0.6]{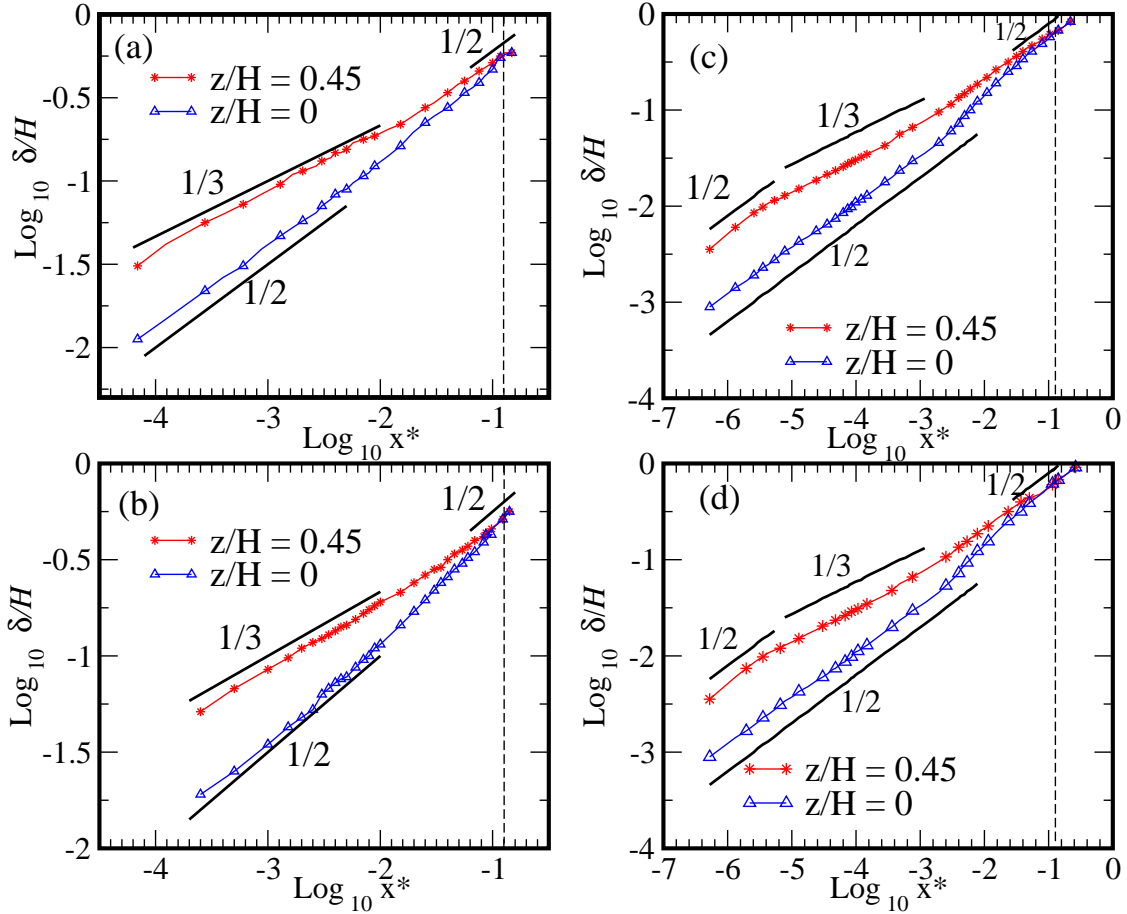}
\caption{Plot of $\log(\delta/H)$ versus $\log(x^{*})$  for aspect ratios of 2 (upper panels) and 5 (lower panels) at \Peclet{} numbers of $\Pe=1000$ (left panels) and $\Pe=10000$ (right panels). The local slope of the lines gives the scaling exponents. The $x$-axis is non-dimensionalized via $x^*=x/(H\Pe)$. The channel height is $H=20$ (left panels) and $H=200$ (right panels). The channel widths are chosen such that the aspect ratio is $W/H=2$ in the case of upper panels and $W/H=5$ in the case of lower panels ($W=40$ (top left), $W=400$ (top right), $W=100$ (bottom left) and $W=1000$ (bottom right)). In all the panels shown, the vertical dashed line indicates $x^*=1/8$ as obtained from a dimensional estimate of the distance for vertical homogenization of the solute concentration.}
\label{fig:scalings}
\end{figure}

As discussed in section \ref{sec:method}, $\delta \sim x^{1/2}$ for diffusion in a uniform velocity field. In the case of the inhomogeneous velocity profile considered here, the relation between the extent of the broadening $\delta$ and the axial distance $x$ takes the more general form  $\delta \sim x^{\gamma}$, where $\gamma$ is the exponent characterizing the diffusive broadening. It is shown in Ref.~\cite{Salmon2007} that $\gamma$ in general depends both on $x$ and $z$. In order to examine this property within our simulations, we survey in \myfig{fig:scalings} the change of $\delta$ downstream along the $x$ direction. The local slope of the (log-log) plots gives the scaling exponent $\gamma$.

\begin{figure}
\includegraphics[scale=0.85, angle =-90]{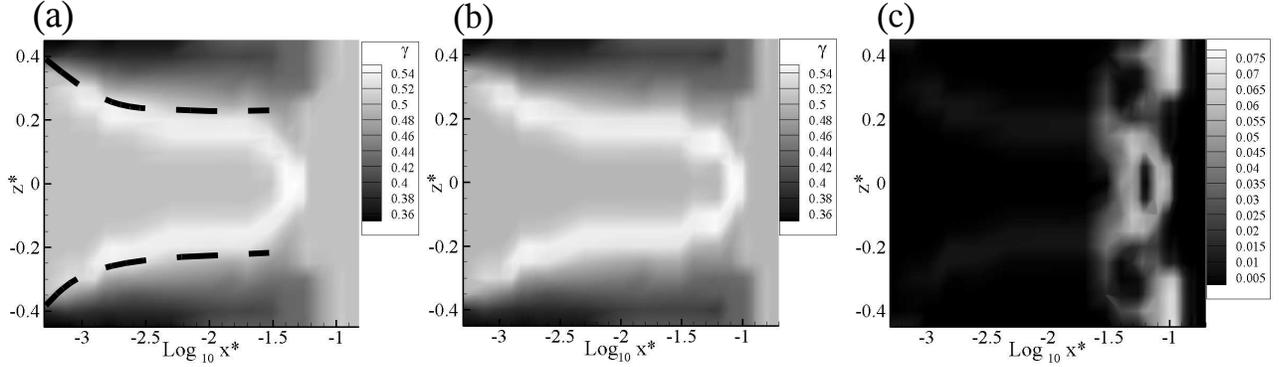}\\[-3mm]
\caption{Plot of the scaling exponent $\gamma$ (appearing in $\delta \propto x^{*\gamma}$) in the $\log(x^*)-z^*$ plane for aspect ratios of (a) 5 and (b) 2. In (c) the  difference, $\gamma$(b)-$\gamma$(a), is shown. The exponent $\gamma$ is obtained by local fits to log-log plot in \myfig{fig:scalings}.}
\label{fig:exponents}
\end{figure}

\begin{figure}
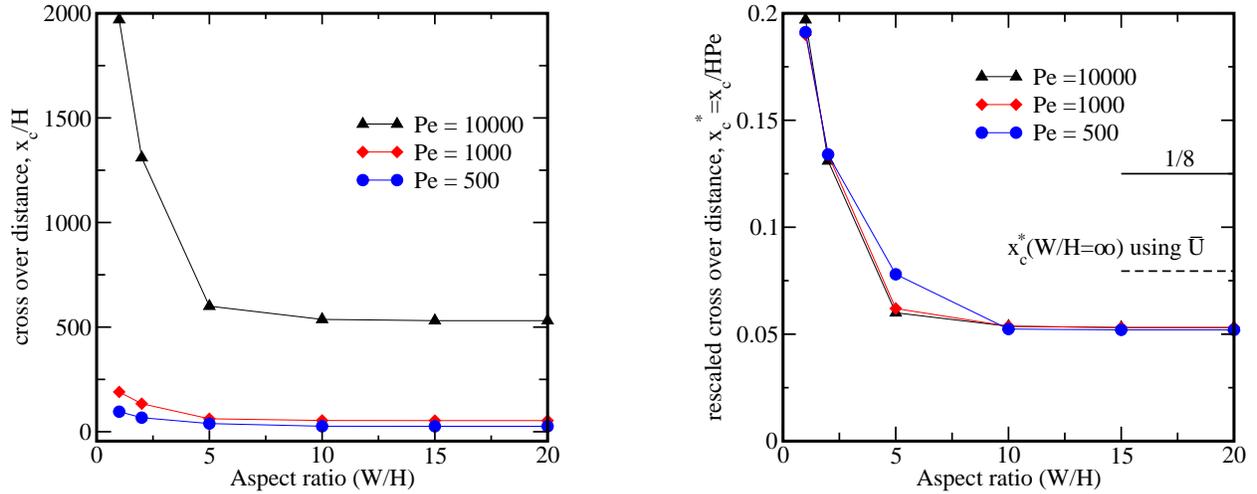

\vspace*{5mm}
\includegraphics[height=65mm]{fig7a.eps}\hfill
\includegraphics[height=65mm]{fig7b.eps}
\caption{Left: Plot of the cross over point $\xc$ versus the aspect ratio for different \Peclet{} numbers as indicated. Right: The same data as in the left panel, rescaled via $\xc^*=\xc /(H\Pe)$. The horizontal solid line marks the value of $\xc^*=1/8$ obtained from an estimate of the time necessary for vertical homogenization in a channel with infinite aspect ratio ($W/H\to \infty$): $2D \tc =H^2/4$. Using the cross-over distance $\xc$ and the mid-channel fluid velocity $U$ to estimate the cross-over time, $\tc=\xc/U$, one obtains $\xc=H^2U/(8D)$ and hence $\xc^*=1/8$ (recall our definition of the \Peclet{} number $\Pe=HU/D$). This estimate is improved significantly by taking, $\tc=\xc/\bar{U}$, where $\bar{U}$ is the average fluid velocity across the channel (dashed horizontal line).\\\\}
\label{fig:xc}
\end{figure}

\begin{figure}
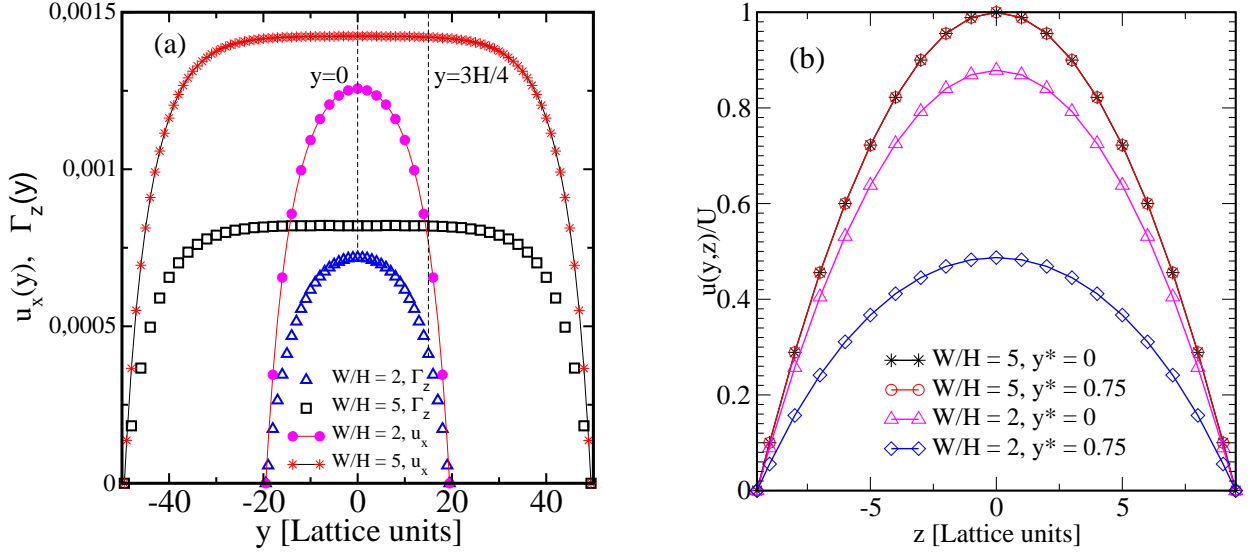

\includegraphics[height=73mm]{fig6a.eps}\hfill
\includegraphics[height=73mm]{fig6b.eps}
\caption{
(a) Velocity, $u_x$, and shear rate, $\Gamma_z\equiv\partial u_x/\partial z$, versus $y$ for aspect ratios of $W/H=2$ and 5. The distance from the bottom wall is equal to $z=H/10$ (=2 lattice units) for all the data shown. For the aspect ratio of 5, there is a wide range of $y$-values for which both $u_x$ and $\Gamma_z$ are roughly constant thus justifying the assumption $u_x(y,z)\simeq u_x(z)$. This assumption, however, fails at the smaller aspect ratio shown. In this case, there is practically no $y$ independent region. Vertical dashed lines mark $y=0$ and $y=3H/4$ for which velocity profiles along the vertical ($z$) direction are depicted in the adjacent panel. (b) Velocity versus $z$ for $y = 0$ and $y=3H/4$ for aspect ratios of $W/H=2$ and 5. In the case of an aspect ratio of $5$, the velocity profile is identical for both values of $y$. This is in accordance with the panel (a) where $u_x$ hardly varies in the $y$ range delimited by the two vertical dashed lines. At a lower aspect ratio of 2, on the other hand, the effect of the side wall is quite significant. The \Peclet{} number is $\Pe=1000$ in all the cases shown.}
\label{fig:Conc_and_Vprofile}
\end{figure}

Within each panel in \myfig{fig:scalings}, the extent $\delta$ of the interdiffusion zone is shown versus axial distance both for the center of the channel ($z/H=0$) as well as in the proximity of one of the walls $(z/H=0.45)$.  The following features can be observed in the two left panels shown in \myfig{fig:scalings} corresponding to a \Peclet{} number of $\Pe=1000$. At small distances from the inlet, $\delta \sim {x}^{1/2}$ at the center of the channel while $\delta \sim {x}^{1/3}$ close to the wall. However, increasing the lattice resolution of our simulation domain in the $z$ direction and correspondingly the \Peclet{} number reveals another short regime, where $\delta \sim {x}^{1/2}$ across the entire cross section of the channel. This regime, as shown in \myfig{fig:scalings} for \Peclet{} number of $\Pe=10000$, occurs earlier in the flow at the entrance of the channel and changes to the 1/3 regime over a short time interval.

The transition of this 1/2 regime to the 1/3 regime can be understood using an analytical argument similar to L\'{e}v\^{e}que analysis by assuming a linear velocity profile very close to the top/bottom wall \cite{Salmon2007}.   Qualitatively, this cross-over represents the enhancement in diffusive broadening arising from a homogeneous shear rate. Since, in the case of the Poiseuille flow studied here, the shear rate is practically zero in the center of the channel, the effect appears only close to the walls, where the velocity profile is approximately linear.

At larger distances from the inlet, on the other hand, the scaling exponent close to the wall gradually increases towards the value of the exponent at the center of the channel, the latter being very close to $1/2$. This latter behavior can be understood by assuming that the cross over from an exponent of $\gamma=1/3$ to $\gamma=1/2$ corresponds to a homogeneous tracer distribution along the $z$-axis. A criterion for vertical homogenization via diffusion is obtained from  $2D \tc = (H/2)^2$ which, using the maximum fluid velocity $U$ to estimate the cross over time, $\tc=\xc/U$, yields the cross over distance $\xc=H^2U/(8D)=H\Pe/8$. In terms of the reduced distance, this relation translates to $\xc^*=1/8$ ~\cite{Salmon2007}.

The above estimate of $\xc$ does not take into account the effect of aspect ratio. In fact, as discussed above, the vertical homogenization takes place at larger axial distances when the aspect ratio is decreased (compare panels (f) and (c) in \myfig{fig:Images}). In order to investigate this issue, data in \myfig{fig:scalings} are plotted for two different aspect ratios of $W/H=2$ (upper panels) and $W/H=5$ (lower panels). Indeed, a comparison of the panels (a) and (b) [as well as (c) and (d)] in \myfig{fig:scalings} suggests that $\xc$ increases when the aspect ratio is decreased. The left and right panels in \myfig{fig:scalings} differ in \Peclet{} number investigated. This is to underline the fact that the observed trend with regard to aspect ratio is not related to the specific choice of the \Peclet{} number.

In order to quantify the effect of aspect ratio further, we determine for two aspect ratios of $W/H=2$ and $W/H=5$ the values of the exponent $\gamma$ for each position $x$ and $z$ along the channel by performing a running fit on the curves in \myfig{fig:scalings} as proposed in~\cite{Salmon2007}. The results are shown in \myfig{fig:exponents}. The dotted black lines in \myfig{fig:exponents}(a) indicate concentration boundary layer which grows with the axial distance at a rate $z^*\sim {x^*}^{1/3}$ \cite{Salmon2007}. The growth of the concentration boundary layer results from the diffusive flux of solute from the top/bottom wall to the center of the channel. The white area of the plot corresponds to the high exponent region of the channel. Our resolution is limited by the small number of points in the $z^{*}$ direction. The difference between the exponents obtained for the two investigated aspect ratios is shown in \myfig{fig:exponents}(c) confirming the retarded cross over to normal diffusive behavior in the case of smaller aspect ratio.

In a more systematic study, we varied the aspect ratio from 1 to 20 and determined the cross over distance $\xc$. The left panel of \myfig{fig:xc} illustrates the behavior of the thus obtained cross over point within the range of aspect ratios investigated and for different \Peclet{} numbers of $\Pe=500,\; 1000$ and $10000$. In line with the results presented above, we observe an increase in the cross over point as the aspect ratio decreases. The right panel of  \myfig{fig:xc} depicts a non-dimensionalized version of the same data.
Interestingly, the data for different \Peclet{} numbers collapse onto a master curve in the limit of high aspect ratios. This is an important observation, since it suggests that, in this limit, the effect of \Peclet{} number is indeed a mere rescaling of time or axial distance. 

In an attempt to better understand the reason for dependence of $\xc$ on aspect ratio, we plot in \myfig{fig:Conc_and_Vprofile}(a) fluid velocity $u_x$ as well as its spatial derivative, $\Gamma_z\equiv\partial u_x/\partial z$, as a function of the transverse coordinate $y$ for aspect ratios of $W/H=2$ and $W/H=5$. The plot is done for a distance of 2 lattice units from the bottom wall. For the aspect ratio of 5, there is a wide range of $y$-values for which both $u_x$ and $\Gamma_z$ are roughly constant thus justifying the assumption that $u_x(y,z)\equiv u_x(z)$ (independent of $y$). This assumption, however, fails at the smaller aspect ratio shown. The data shown in the right panel of \myfig{fig:Conc_and_Vprofile} further underline the importance of taking into account the dependence of the fluid velocity both on $z$ and on $y$ when the aspect ratio is small.

The data shown in \myfig{fig:Conc_and_Vprofile} suggest that the increase in $\xc$ is probably due to the side wall shear effect on solute distribution. At a lower aspect ratio, the shear rate $\Gamma_{z}$ decreases strongly close to the side wall. An estimate from \myfig{fig:Conc_and_Vprofile}(a) shows that for an aspect ratio of 2, the $\Gamma_{z}$ at position $y = 3H/4$  is about 40\% less than that at the center of the channel ($y=0$). Therefore, solute diffusing towards the side walls samples a strongly decreasing shear rate. Given that the extent of broadening $\delta$ at the top/bottom wall takes the form $\delta \sim (xD/\Gamma_{z})^{1/3}$ \cite{Ismagilov2000},  the interdiffusion becomes faster when approaching the side walls, whereby enhancing the inhomogeneity of diffusion (the so called ``butterfly effect'') which is already present at infinite aspect ratio. This is exactly what we observe in the concentration profile images shown in \myfig{fig:Images}.

  In general, solute diffusing towards the side walls from the center of the channel samples constant velocity gradient $\Gamma_{z}$ up to a distance $H$ from the side wall. The time spent by solute before sampling a substantial decrease in the velocity gradient is, therefore,  $t_{\mathrm{W}}  \sim (W/2-H)^2/2D$. For $W \gg H$, this time scale is greater than the time scale to diffuse from the top/bottom wall to the center of the channel denoted as $t_{\mathrm{H}} \sim (H/2)^2/2D$. Thus, at high aspect ratio, solute concentration becomes homogeneous along the vertical direction long before the side walls are ``felt''. Consequently, the cross over point $\xc$ becomes independent of aspect ratio. In the case where $t_{\mathrm{W}} \le t_{\mathrm{H}}$, on the other hand, one can not neglect the additional enhancement of inhomogeneity of diffusive broadening due to the side wall effect. This leads to $W \le 3H$ for the effect of aspect ratio being significant. As a survey of $\xc^*$ in \myfig{fig:xc}(b) reveals, this simple estimate (which is based on a dimensional argument only) lies within a factor of two of the result obtained within our computer simulations.

\subsection{Summary}
 We study the effect of a finite aspect ratio on transverse diffusive transport of miscible solutes flowing in a pressure driven microchannel using the lattice Boltzmann method. The lattice Boltzmann method incorporates the essential 3D features such as the non-parabolicity of the velocity profile and the velocity gradient due to the side walls. We observe the previously reported \cite{Ismagilov2000,Kalmholz2002,Salmon2007,Jimenez2005} different exponents characterizing the extent of the broadening both at the early stage of mixing of the two fluids and at the later stage downstream. Interestingly, we observe the same scaling laws regardless of the channel aspect ratio. However, extent of diffusive broadening and the position, $\xc$, at which the broadening becomes uniform and finally reverts to the 1/2 behavior vary remarkably with the  channel aspect ratio. The \Peclet{} number, on the other hand, is found to play the role of a scale factor in $\xc$ in a way that $\xc^* \equiv \xc/(H\Pe)$ is independent of aspect ratio. This is inline with the general structure of the advection-diffusion equation, upon neglection of axial diffusion.

 A qualitative understanding of the effect of aspect ratio is provided invoking the influence of the shear stress in the proximity of the side walls on diffusive broadening. This is based on the idea that side wall shear stress is non-negligible in a region of width $H$ close to the side walls. The corresponding effect on inhomogeneous diffusive broadening will be felt if the solutes at the center of the channel have enough time to reach this region {\it before} vertical homogenization takes place. This allows to derive a simple criterion to decide whether a given aspect ratio is  ``large'' or ``small''. Within a factor of two, this estimate correctly reproduces the behavior observed within our lattice Boltzmann computer simulations.

\acknowledgements{This work was supported by the Max-Planck Initiative for Multiscale Materials Modeling of Condensed matter (MMM).}


\begin{thebibliography}{9}
	\bibitem{Lipman2003}
	E.A. Lipman, B. Schuler, O. Bakajin, and W.A. Eaton, Science, {\bf 301}, 1233 (2003).
        \bibitem{Seong2003}
        G. H. Seong, Jinseok Heo, and R.M. Crooks, Anal. Chem., {\bf 75}, 3161 (2003).
        \bibitem{David2002}
        J.B. David, A.M. Glennys, and G.M. Walker Annu Rev. Biomed. Eng. {\bf 4}, 261(2002).
        \bibitem{Kalmholz1999}
        A.E. Kalmholz, B.H. Weigl, B.A. Finlayson, and P. Yager, Anal. Chem. {\bf 71}, 5340 (1999).
        \bibitem{Ismagilov2000}
        R.F. Ismagilov, A.D. Strock, P.J.A. Kennis, G. Whitesides, and H.A. Stone,
        Appl. Phys. Lett. {\bf 76}, 2376 (2000).
        \bibitem{Leveque1928}
        M.A. L\'{e}v\^{e}que, Ann. Mines, {\bf 13}, 201 (1928).
        \bibitem{Kalmholz2002}
        A.E. Kalmholz, and P. Yager, Sensors, Actuators B {\bf 82}, 117 (2002).
        \bibitem{Salmon2007}
        Jean-Baptiste Salmon, and A. Adjari, J. Appl. Phys. {\bf 101}, 074902 (2007).
        \bibitem{Jimenez2005}
        J. Jim\'{e}nez J. Fluid Mech. {\bf 535}, 245 (2005).
        \bibitem{Gondret1997}
        P. Gondret, N. Rakotomalala, M. Rabaud, D. Salin, and P. Watzky, Phys. Fluids {\bf 9}, 1841 (1997).
        \bibitem{Zheng2008} 
        Xu Zheng, and Zhan-hua Silber-Li, Exp. Fluids {\bf 44}, 951 (2008).
        \bibitem{Qian1992}
         Y.H. Qian, D. d'Humi\`eres, and P. Lallemand, Euro. Phys. Lett. {\bf 17}, 479 (1992).
        \bibitem{Varnik2006}
        F. Varnik, and D. Raabe, Modelling Simul. Mater. Sci. Eng. {\bf 14}, 857 (2006).
       \bibitem{Varnik2007a}
        F. Varnik, D. Dorner, and D. Raabe, J.~Fluid Mech. {\bf 573}, 191 (2007).
        \bibitem{Varnik2007b}
        F. Varnik, and D. Raabe, Molecular Simulation {\bf 33}, 583 (2007).
       \bibitem{McNamara88}
        G. McNamara and G. Zanetti, Phys. Rev. Lett. {\bf 61},  2332  (1988).
       \bibitem{Higuera89}
        F. Higuera, S. Succi, and R. Benzi, Europhys. Lett. {\bf 9},  345  (1989).
       \bibitem{Qian92}
        Y. Qian, D. d'Humieres, and P. Lallemand, Europhys. Lett. {\bf 17},  479  (1992).
        \bibitem{succi2001}
        S. Succi, The lattice Boltzmann Equation: for Fluid Dynamics and Beyond, Oxford University Press, (2001).
        \bibitem{He1997}
        X. He, and L.S. Luo, Phys.~Rev.~E {\bf 56}, 6811 (1997).
        \bibitem{Rubinstein2008}
         R. Rubinstein, and L.S. Luo, Phys. Rev. E {\bf 77}, 036709 (2008).
        \bibitem{Frisch1987}
         U. Frisch, D. d'Huim\`{e}res, B. Hasslacher, P. Lallemand, Y. Pomeau, and J.-P. Rivet, Complex Systems {\bf 1}, 649, (1987).
         \bibitem{Koelman1991}
         J.M.V.A. Koelmann, Euro.~Phys.~Lett {\bf 15}, 603 (1991).
         \bibitem{Schlichting1958}
         H. Schlichting, Boundary-Layer Theory, McGraw Hill, London (1958)
         \bibitem{Mathews1970}
        J. Mathews, and R.L. Walker, Mathematical methods of Physics, Second Ed. Addison-Wesley (1970).
         \bibitem{Crank1956}
         J. Crank, The mathematics of diffusion, Oxford, London (1956).
        

\end{thebibliography}
\end{document}